# Atomic-scale Electronic Structure of the Cuprate *d*-Symmetry Form Factor Density Wave State

M. H. Hamidian[1†], S.D. Edkins[1,2†], Chung Koo Kim[3], J. C. Davis[1,2,3,4], A. P. Mackenzie[2,5], H. Eisaki[6], S. Uchida[7], M. J. Lawler[1,8], E.-A. Kim[1], S. Sachdev[9,10], and K. Fujita[3]

1. LASSP, Department of Physics, Cornell University, Ithaca, NY 14853, USA.
2. School of Physics and Astronomy, University of St. Andrews, Fife KY16 9SS, Scotland.
3. CMPMS Department, Brookhaven National Laboratory, Upton, NY 11973, USA.
4. Kavli Institute at Cornell for Nanoscale Science, Cornell University, Ithaca, NY 14853, USA.
5. Max-Planck Institute for Chemical Physics of Solids, D-01187 Dresden, Germany.
6. Institute of Advanced Industrial Science and Technology, Tsukuba, Ibaraki 305-8568, Japan.
7. Department of Physics, University of Tokyo, Bunkyo-ku, Tokyo 113-0033, Japan.
8. Dept. of Physics and Astronomy, Binghamton University, Binghamton, NY 13902.
9. Department of Physics, Harvard University, Cambridge, MA.
10. Perimeter Institute for Theoretical Physics, Waterloo, Ontario N2L 2Y5, Canada
† These authors contributed equally to this project.

ABSTRACT

Extensive research into high temperature superconducting cuprates is now focused upon identifying the relationship between the classic 'pseudogap' phenomenon[1,2] and the more recently investigated density wave state[3-13]. This state always exhibits wavevector $Q$ parallel to the planar Cu-O-Cu bonds[4-13] along with a predominantly *d*-symmetry form factor [14-17] (dFF-DW). Finding its microscopic mechanism has now become a key objective[18-30] of this field. To accomplish this, one must identify the momentum-space (*k*-space) states contributing to the dFF-DW spectral weight, determine their particle-hole phase relationship about the Fermi energy, establish whether they exhibit a characteristic energy gap, and understand the evolution of all these phenomena throughout the phase diagram. Here we use energy-resolved sublattice visualization[14] of electronic structure and show that the characteristic energy of the dFF-DW modulations is actually the 'pseudogap' energy $\Delta_1$. Moreover, we demonstrate that the dFF-DW modulations at $E=-\Delta_1$ (filled states) occur with relative phase π compared to those at E=$\Delta_1$ (empty states). Finally, we show that the dFF-DW $Q$ corresponds directly to scattering between the 'hot frontier' regions of $k$-space beyond which Bogoliubov quasiparticles cease to exist[31,32,33]. These data demonstrate that the dFF-DW state is consistent with particle-hole interactions focused at the pseudogap energy scale and between the four pairs of 'hot frontier' regions in $k$-space where the pseudogap opens.

A conventional 'Peierls' charge density wave (CDW) in a metal results from particle-hole interactions which open an energy gap at specific regions of $\boldsymbol{k}$-space that are connected by a common wavevector $\boldsymbol{Q}$. This generates a modulation in the density of free charge at $\boldsymbol{Q}$ along with an associated modulation of the crystal lattice parameters. Such CDW states are now very well known[34]. In principle, a density wave modulating at $\boldsymbol{Q}$ can also exhibit a 'form factor' (FF) with different possible symmetries[35,36] (see Supplementary Information (SI) Section 1). This is relevant to the high-temperature superconducting cuprates because numerous researchers have recently proposed that the 'pseudogap' regime[1,2] (PG in Fig. 1A) contains an unconventional density wave with a $d$-symmetry form factor[18-30]. The basic phenomenology of such a state is that intra-unit-cell (IUC) symmetry breaking renders the $O_x$ and $O_y$ sites within each CuO$_2$ unit-cell electronically inequivalent, and that this inequivalence is then modulated periodically at wavevector $\boldsymbol{Q}$ parallel to (1,0);(0,1). The real-space ($\boldsymbol{r}$-space) schematic of such a $d$-symmetry FF density wave (dFF-DW) at $\boldsymbol{Q_x}$ as shown in Fig. 1B exemplifies periodic modulations at the $O_x$ sites that are π out of phase with those at the $O_y$ sites. This state is then described by $A(\boldsymbol{r}) = D(\boldsymbol{r})Cos(\phi(\boldsymbol{r}) + \phi_0(\boldsymbol{r}))$, where $A(\boldsymbol{r})$ represents the modulating electronic degree of freedom, $\phi(\boldsymbol{r}) = \boldsymbol{Q_x} \cdot \boldsymbol{r}$ is the DW spatial phase at location $\boldsymbol{r}$, $\phi_0(\boldsymbol{r})$ represents disorder related spatial phase shifts, and $D(\boldsymbol{r})$ is the magnitude of the $d$-symmetry form factor[14,22,24]. To distinguish between the various microscopic mechanisms proposed for the $\boldsymbol{Q}$=(Q,0);(0,Q) dFF-DW state of cuprates[18-30], it is essential to establish its atomic-scale phenomenology, including the **k**-space eigenstates contributing to its spectral weight, the relationship (if any) between modulations occurring above and below the Fermi energy, whether the modulating states in the DW are associated with a characteristic energy gap, and how the dFF-DW evolves with doping.

To visualize such phenomena directly as in Fig. 1C, we use sublattice phase-resolved imaging of the electronic structure[14] of the $CuO_2$ plane. Both the scanning tunneling microscope (STM) tip-sample differential tunneling conductance $\frac{dI}{dV}(r, E = eV) \equiv g(r, E)$ and the tunnel-current $I(r, E)$ are measured at bias voltage $V=E/e$ and with sub-unit-cell spatial resolution. Because the density-of-electronic-states $N(r, E)$ is related to the differential conductance as $g(r, E) \propto \left[eI_s / \int_0^{eV_s} N(r, E')dE'\right] N(r, E)$ with $I_s$ and $V_s$ being arbitrary parameters and the denominator $\int_0^{eV_s} N(r, E')dE'$ unknown, valid determination of $N(r, E)$ is impossible (SI Section 2). However, one can suppress these serious systematic errors by using $R(r, E) = I(r, E)/I(r, -E)$ or $Z(r, E) = g(r, E)/g(r, -E)$; this allows distances, wavelengths, and spatial-phases of electronic structure to be measured accurately. The unprocessed $g(r, E)$ acquired for and analyzed in this paper: (i) are measured over very large fields of view (to achieve high phase-precision in Fourier analysis) ; (ii) simultaneously maintain deeply sub-unit-cell precision measurements in **r** (to achieve high precision in sublattice segregation); (iii) are taken over a wide range of energies $E$ with fine energy-spacing, so that energy dependences of $d$-symmetry FF modulations may be accurately determined. We then calculate each sublattice-phase-resolved $Z(r, E)$ image and separate it into three: the first, $Cu(\mathbf{r})$, contains only the measured values of $Z(\mathbf{r})$ at Cu sites while the other two, $O_x(\mathbf{r})$ and $O_y(\mathbf{r})$, contain only the measurements at the $x/y$-axis oxygen sites. Phase-resolved Fourier transforms of the $O_x(\mathbf{r})$ and $O_y(\mathbf{r})$ sublattice images[14],

$\tilde{O}_x(q) = Re\tilde{O}_x(q) + iIm\tilde{O}_x(q); \tilde{O}_y(q) = Re\tilde{O}_y(q) + iIm\tilde{O}_y(q)$, are used to determine the form factor symmetry for modulations at any $q$

$$\tilde{D}^Z(q) = (\tilde{O}_x(q) - \tilde{O}_y(q))/2 \qquad (1a)$$
$$\tilde{S}'^Z(q) = (\tilde{O}_x(q) + \tilde{O}_y(q))/2 \qquad (1b)$$
$$\tilde{S}^Z(q) = \widetilde{Cu}(q) \qquad (1c)$$

where the superscript Z identifies the type of sublattice resolved data used. Specifically for a DW occurring at $Q$, one can then evaluate the magnitude of its $d$-symmetry form factor $\tilde{D}(Q)$ and its $s'$- and $s$-symmetry form factors $\tilde{S}'(Q)$ and $\tilde{S}(Q)$, respectively. Studies of non-energy-resolved $R(r,E)$ images using this approach have revealed that the DW modulations in the $O_x(r)$ and $O_y(r)$ sublattice images of electronic structure in underdoped $Bi_2Sr_2CaCu_2O_{8+x}$ and $Ca_{2-x}Na_xCuO_2Cl_2$ consistently exhibit a relative phase of π and therefore a $d$-symmetry form factor[14]; X-ray scattering studies[15,16] yield the same conclusion for two other cuprates, $YBa_2Cu_3O_{7-x}$ and $Bi_2Sr_{2-x}La_xCuO_{6+δ}$.

X-ray scattering studies report a short-ranged density wave with wavevector centered around $Q$=(Q,0);(0,Q) occurring approximately in the pink shaded regions[11,12,13] of the schematic phase diagram in Fig 1A. Figures 1B,C exemplify the predominately $d$-symmetry form factor[14-17] of this DW when imaged directly. One obstacle to understanding this dFF-DW state is that large field-of-view sublattice-resolved images of cuprate electronic structure[14] never look like an ideal long-range ordered version of Fig. 1B. Instead, Fig. 2A shows a typical $Z(r,150meV)$ image of $p$=8% $Bi_2Sr_2CaCu_2O_{8+x}$, for $T<<T_c$ in the superconducting phase while Fig. 2B shows the equivalent $Z(r,150meV)$ for $T>T_c$ in the cuprate pseudogap phase. While elements indistinguishable from Fig. 1C can be seen in 2A, B, no long range order is obvious. Therefore, to explore the spatial arrangements of the dFF-DW in such electronic structure images, we analyze $\tilde{D}^Z(q)$ which is a robust sublattice-phase-resolved measure of the $d$-symmetry form factor (SI Section 3). Analysis of Figures 2A, B in this fashion yields 2C,D; both clearly exhibit the dFF-DW maxima at the two inequivalent wavevectors $\boldsymbol{Q}_x$ and $\boldsymbol{Q}_y$. Fourier filtering these two $\tilde{D}^Z(\boldsymbol{q})$ from 2A,B for only those regions surrounding $\boldsymbol{Q}_x$ and $\boldsymbol{Q}_y$ (within dashed circles) generates two complex-valued images $D_x(\boldsymbol{r}), D_y(\boldsymbol{r})$

$$D_x(\boldsymbol{r}) = \frac{2}{(2\pi)^2} \int d\boldsymbol{q}\, e^{i\boldsymbol{q}\cdot\boldsymbol{r}} \tilde{D}^Z(\boldsymbol{q}) e^{-\frac{(\boldsymbol{q}-\boldsymbol{Q}_x)^2}{2\Lambda^2}} \quad;\quad D_y(\boldsymbol{r}) = \frac{2}{(2\pi)^2} \int d\boldsymbol{q}\, e^{i\boldsymbol{q}\cdot\boldsymbol{r}} \tilde{D}^Z(\boldsymbol{q}) e^{-\frac{(\boldsymbol{q}-\boldsymbol{Q}_y)^2}{2\Lambda^2}} \qquad (2a)$$

where $\Lambda^{-1}$ is the characteristic length scale over which variations in $D_x(\boldsymbol{r}), D_y(\boldsymbol{r})$ can be resolved, set by the filter width in Fourier space.

Their magnitudes

$$|D_x(\boldsymbol{r})| = \sqrt{\left(ReD_x(\boldsymbol{r})\right)^2 + \left(ImD_x(\boldsymbol{r})\right)^2} \qquad (2b)$$

$$|D_y(\boldsymbol{r})| = \sqrt{\left(ReD_y(\boldsymbol{r})\right)^2 + \left(ImD_y(\boldsymbol{r})\right)^2} \qquad (2c)$$

represent the local amplitudes of dFF-DW modulations along $\boldsymbol{Q}_x$ and $\boldsymbol{Q}_y$, respectively. Any unidirectional domain arrangements of the dFF-DW state can then be determined by introducing

$$F(\boldsymbol{r}) = \frac{|D_x(\boldsymbol{r})| - |D_y(\boldsymbol{r})|}{|D_x(\boldsymbol{r})| + |D_y(\boldsymbol{r})|} \qquad (3)$$

which is designed to identify regions where the dFF-DW modulation is primarily along the *x*-axis or the *y*-axis depending on the sign of $F(\boldsymbol{r})$ (SI Section 4). Figures 2E,F show how regions of $-1.0 < F(\boldsymbol{r}) < -0.3$ (shaded blue) are primarily modulating along *y*-axis while regions $+0.3 < F(\boldsymbol{r}) < +1.0$ (shaded orange) are primarily modulating along *x*-axis (those with $-0.3 < F(\boldsymbol{r}) < +0.3$ shaded white appear at boundaries). Figures 2G and 2H reveal the results of this analysis for the data in Figs 2A,B respectively. Overall, the system is configured into spatial regions within which the dFF-DW along one direction is dominant. By overlaying the color scale for $F(\boldsymbol{r})$ on the data in Figs 2A,B to create Figs 2G,H, one can see directly the unidirectional region configurations derived from Eqn. 3.

These observations of coexisting nanoscale unidirectional regions are in reasonable agreement with related X-ray studies[37] of $YBa_2Cu_3O_{7-x}$. Finally, since the data in Fig. 2B and Fig. 2H were measured at $T>T_c$ (pink region Fig. 1A), it demonstrates directly that the cuprate dFF-DW appears first in the non-superconducting 'pseudogap' regime.

A conventional CDW state opens a gap in the energy spectrum of ***k***-space electronic eigenstates with the maximum spectral weight of modulating states occurring at the edges of this energy-gap[34]. But which energy gap (if any) is associated with the dFF-DW state found in underdoped cuprates is unknown. Figure 3A shows how a typical tunneling conductance spectrum representative of strongly underdoped cuprates exhibits two characteristic energies [31,32,33]. While the lower energy scale $\Delta_0$ represent the maximum energy at which Bogoliubov quasiparticle excitations exist[31,32,33] (see Figure 3B) the higher energy scale (dashed blue line) is the cuprate 'pseudogap' as determined from its comparison with doping dependence of pseudogap scale in tunneling and photoemission. To identify the energy dependence of the cuprate dFF-DW states, we measure $Z(\bm{r},|E|)$ and from it calculate $\tilde{D}^Z(\bm{q},|E|)$, $\tilde{S}'^Z(\bm{q},|E|)$, and $\tilde{S}^Z(\bm{q},|E|)$. Figure 3C shows the measured power-spectral-density of the *d*-symmetry FF modulations $\tilde{D}^Z(\bm{q},|88\text{meV}|)^2$, with the wavevectors near $\bm{Q}_x$ and $\bm{Q}_y$ indicated by red circles. Adopting the common convention in X-ray studies[9,10,11,16] for determining the DW wavevector magnitude $|\bm{Q}|$, we measure $\tilde{D}^Z(\bm{q},|E|)$ along a line in the high symmetry directions (1,0):(0,1) passing through the region of the dFF-DW peak and fit these data to a background plus Gaussian. The peak positions of the two Gaussians are then assigned to be the values of $\bm{Q}_x$ and $\bm{Q}_y$ (although a broad range of values of $\bm{Q}$ can actually be detected under each peak in $\tilde{D}^Z(\bm{q},|E|)$ e.g. dashed circles Figs 2C,D). It remains to be determined whether these incommensurate maxima in $\tilde{D}^Z(\bm{q},|E|)$ are due to domains of continuously incommensurate dFF-DW or domains of commensurate dFF-DW separated by discommensurations[38]. Nevertheless, Figure 3B plots the energy dependence of the dFF-DW wavevectors (blue line) determined in this way for a $p = 0.06$ sample. Such information was not previously available from measurements of the modulation wavevectors from STM images lacking sublattice phase-resolved segregation[14] into $\tilde{O}_x(\boldsymbol{q})$ and $\tilde{O}_y(\boldsymbol{q})$. Figure 3E shows the measured $\boldsymbol{k}$-space locus where Bogoliubov quasiparticles exist[31,32,33] as a function of hole density. When the dispersive 'octet' of Bogoliubov scattering interference disappears a transition occurs to a non-dispersive density wave modulation (Fig. 3B). In Fig. 3D, the doping dependence of the conventional $\boldsymbol{Q}_x$, $\boldsymbol{Q}_y$ of the $d$-symmetry form factor modulations is shown using blue symbols, while the measured shortest wavevectors interconnecting the $\boldsymbol{k}$-space arc tips (Fig. 3E) are indicated by using colored symbols referring to each hole-density in Fig. 3E. These data demonstrate directly that the conventional $\boldsymbol{Q}_x$, $\boldsymbol{Q}_y$ of the dFF-DW state corresponds to the locations in $\boldsymbol{k}$-space of arc tips. Finally, in Fig. 3F we show the measured energy dependences of the amplitudes of the s, s'- and $d$-form factor modulations, $S^Z(E)$, $S'^Z(E)$ and $D^Z(E)$, determined by integrating over the region of $\boldsymbol{q}$-space enclosed by solid red circles in 3C (SI Section 5). The $d$-symmetry form factor is negligible for modulations in the low energy range that contains only Bogoliubov quasiparticles (and which we now see is dominated by s'-symmetry form factors) but it rapidly becomes intense at higher energy and reaches maximum at the pseudogap energy scale which for this sample is $\Delta_1 \sim 90$meV. This reveals that the characteristic energy of electronic-structure modulations in the cuprate $d$-symmetry FF density wave is actually the pseudogap energy.

As a function of energy, the transition from Bogoliubov quasiparticle interference modulations to dFF density wave modulations occurs in an unusual fashion. Although Bogoliubov QPI is observed as expected everywhere on the Fermi surface in overdoped cuprates[33], in underdoped samples it evolves as expected only until the energy $E \approx \Delta_0$ at which the terminations of the Bogoliubov coherent $\boldsymbol{k}$-space arcs (Fig. 3E) are observed[31,32,33]. Here, the set of seven dispersive scattering interference modulations $q_1, q_2, ... q_7$ signifying Bogoliubons[32] (SI Section 5) disappears in a narrow energy window during which dispersion of the two surviving modulations q$_1$(E) and q$_5$(E) comes to a halt, leaving the non-dispersive dFF-DW modulations with $q_1^* \approx q_1(\Delta_0)$ and $q_5^* \approx q_5(\Delta_0)$ (see Fig. 3B and Refs. 31,32,33). The intensity of these modulations first becomes detectable at $\Delta_0$ and, as we show below, reaches an intense maximum at $\Delta_1$, all the while maintaining the same wavevectors $Q_x^d$ and $Q_y^d$ as shown in Fig. 3B. We refer to this *k*-space region where Bogoliubov quasiparticles yield to modulations of a dFF-DW as the 'hot frontier'[39] to distinguish it from the colloquial 'hot spots' beyond which, in a conventional density wave, dispersive quasiparticle states would reappear. In cuprates, this does not occur and, instead, the 'hot frontiers' define the *k*-space limit beyond which only dFF-DW modulations are detected[31,32,33] using SI-STM (blue in Fig. 3B).

Key information on the microscopic cause of any DW state is also contained in the relationship between modulations of states above and below the Fermi energy. For example, mixing via interactions of states with momenta $\boldsymbol{k_1}$ and $\boldsymbol{k_2}$ generates modulations at wavevector $\boldsymbol{Q} = \boldsymbol{k_1} - \boldsymbol{k_2}$. The wavefunctions of any resulting DW would then form bonding/anti-bonding states below/above the Fermi level which are proportional to $e^{i\boldsymbol{k_1}\cdot\boldsymbol{r}} \pm e^{i\boldsymbol{k_2}\cdot\boldsymbol{r}}$. The related densities of these states would then exhibit modulations governed by $\left|e^{i\boldsymbol{k_1}\cdot\boldsymbol{r}} \pm e^{i\boldsymbol{k_2}\cdot\boldsymbol{r}}\right|^2 = 2(1 \pm Cos(\boldsymbol{Q}\cdot\boldsymbol{r}))$. In such scenarios, the DW modulations above the Fermi energy should always be $\pi$ out-of-phase with the equivalent ones below. To explore this issue in $Bi_2Sr_2CaCu_2O_{8+x}$, we show in Figs 4A,B the measured $g(\boldsymbol{r}, +94\text{meV})$ from filled states and $g(\boldsymbol{r}, -94\text{meV})$ from empty states, respectively, each at the characteristic energy of the dFF-DW (Fig 3F and SI Section 6). For these two images the sublattice-phase-resolved $\widetilde{D}^g(\boldsymbol{q}, E)$ (Eqn. 1a) are calculated and reveal a predominantly *d*-symmetry form factor modulation with wavevectors near $Q_x$ and $Q_y$ in Figs 4A,B. Next, by Fourier filtering these two $\widetilde{D}^g(q,E)$ for regions surrounding $Q_y$ we determine the complex-valued $D_y(r)$ and thus the spatial phase of dFF-DW modulation along $Q_y$ as

$$\phi_y(r,E) = \arctan\left(ImD_y(r,E)/ReD_y(r,E)\right) \qquad (4a)$$

For $E=+94$meV this is shown in Fig. 4C and for $E=-94$meV in Fig. 4D. Visual comparison reveals that these two $\phi_y(r,\pm E)$ images are out of phase with each other by π. And indeed, the spatial-average value of $\phi_y(r,+E) - \phi_y(r,-E)$ as a function of $E$ (over the whole field of view A and B) is shown in Fig. 4F. It reveals that, while the relevant $Q_x$ and $Q_y$ components of $g(r,+E)$ and $g(r,-E)$ images are in phase with each other at low energy, they rapidly evolve at $|E|>70$meV and become globally π out of phase at $|E|\sim\Delta_1$ (Figs 4A,B). The shaded region indicates evolution through a situation where some areas exhibit $\phi \sim 0$ and some $\phi \sim \pi$ but this is quickly resolved upon reaching pseudogap energy $\Delta_1$. Similar analysis for the particle-hole symmetry in phases $\phi_x(r,\pm E)$ of $Q_x$ modulations

$$\phi_x(r,E) = \arctan\left(ImD_x(r,E)/ReD_x(r,E)\right) \qquad (4b)$$

yields a virtually identical result. These phenomena also occur throughout the underdoped regions the phase diagram (SI Section 6.III). All these data demonstrate that a phase difference of π exists between spatial modulations of the filled states at pseudogap energy $E\sim-\Delta_1$ and the empty states at $E\sim+\Delta_1$, for the cuprate dFF-DW state.

To summarize: by introducing new techniques to determine the energy/momentum and doping dependence of modulation form factor symmetry, we find that the predominantly *d*-symmetry form factor density wave exists throughout the underdoped region of the $Bi_2Sr_2CaCu_2O_8$ phase diagram (Fig. 3D) including in the pseudogap regime $T>T_c$ (Figs 1C, 2B). The spatial arrangements are primarily in the form of nanoscale regions each containing a primarily unidirectional dFF-DW (Figs 2G,H). The conventionally defined wavevectors $Q_x$ and $Q_y$ of the dFF-DW state evolve with doping as determined by the four shortest scattering vectors linking the $k$-space regions beyond which Bogoliubov quasiparticle excitations are nonexistent (Figs 3D,E) and at which the pseudogap emerges. Further, we demonstrate that, as determined in terms of tunneling probabilities, the dFF-DW state is particle–hole antisymmetric in the sense that a phase difference of π exists between spatial modulations of the filled states ($E\sim-\Delta_1$) and the empty states ($E\sim+\Delta_1$) (Fig. 4E). Most significantly, we show that the characteristic energy of the dFF-DW electronic-structure modulation is actually the pseudogap energy $\Delta_1$ (Fig. 3F).

These data provide evidence that the cuprate *d*-symmetry form factor density wave state involves particle-hole interactions, and that these occur close to wavevectors interconnecting the 'hot frontiers' in $k$-space at which the pseudogap emerges[31,32,33]. Moreover, the dFF-DW electronic structure modulations have a characteristic energy scale which is the pseudogap energy. This intimate connection of the dFF-DW state with the pseudogap electronic structure is consistent with the fact that this state is only found within the pseudogap regime[11-13] Of course, electron-lattice interactions can also play a significant role, with the coupling to the $B_{1g}$ modes long being of foremost interest[20,21,40]. Strong interactions of this mode with the electrons[41] ultimately leading to static, finite $Q$, lattice distortions with *d*-symmetry form factor[42] have recently been discovered in association with the cuprate dFF-DW state. Nevertheless, electron-lattice interactions are not by themselves sufficient to explain the phase diagram of the dFF-DW[11-13] because, for example, they also exist in the overdoped regime where the dFF-DW is absent. Moreover, theoretical models involving $k$-space instabilities[27,28,30,43] which are consistent with the results herein, emphasize that a density wave with this $Q$ and form factor symmetry cannot emerge from a large Fermi surface; instead, a preexisting reorganization of $k$-space due to the pseudogap would be required. Overall, our data support a microscopic picture in which the exotic electronic structure of the pseudogap is parent to the $d$FF-DW state and not vice-versa, where the energy-scale and wavevectors of the dFF-DW are intimately linked to those of the pseudogap, and in which the dFF-DW competes directly for spectral weight with the d-symmetry superconductor at the $\boldsymbol{k}$-space 'hot frontier' between superconductivity and the pseudogap.

**Figure Captions**

**Figure 1 *d*-symmetry form factor density wave in cuprate pseudogap phase.**

A. Schematic phase diagram of hole-doped cuprates. The pseudogap regime has been identified by, for example, suppression of uniform magnetic susceptibility and electronic specific heat, and the appearance of a truncated Fermi surface referred to as the 'Fermi Arc' [1,2]. The dome-shaped region of *d*-symmetry Cooper paired high temperature superconductivity is universally accepted. More recently, an unusual density wave state has been detected by bulk probes[4-13] in the region indicated schematically in pink; its modulations are now known to have a *d*-symmetry form factor [14-17]. The range of hole-density, *p*, in which *d*-symmetry form factor density waves are studied in this paper is indicated by the white double-headed arrow.

B. Schematic of the electronic structure in a cuprate dFF-DW. Grey dots represent the Cu sites and correspond to the white dots in 1C. The $O_x$ and $O_y$ sites within each $CuO_2$ unit-cell are electronically inequivalent as represented by a color scale ranging from yellow through white to blue. The schematic DW modulates horizontally with wavelength λ or with wavevector $Q_x$ (horizontally) and with period $4a_0$. The periodic modulations at $O_x$ sites are π out of phase with those at $O_y$ sites as seen by considering the two trajectories marked $\phi_x$ and $\phi_y$ (SI Section 1).

C. Measured $Z(r,150\text{meV})$ at $T>T_c$ in the pseudogap phase of $Bi_2Sr_2CaCu_2O_8$ at hole-density $p\sim8\%$. Two periods of dFF-DW modulation at $Q_x$ that correspond directly to the schematic in 1B are shown. Thus, to observe the dFF-DW state sublattice-phase-resolved imaging is required and achieved here in the pseudogap regime. The transparent overlay shows the relationship between locations of Cu, $O_x$, $O_y$ atoms in the $CuO_2$ plane and the dFF-DW whose wavevector here is along the x-direction

**Figure 2 dFF-DW domains in superconducting and pseudogap phases.**

A. Measured $Z(r,150\text{meV})$ at $T \ll T_c$ in the superconducting phase of hole density $p \sim 8\%$ doped $Bi_2Sr_2CaCu_2O_{8+x}$ ($T \sim 4.2K$). These complex spatial features involve modulations that comprehensively maintain a relative phase of $\pi$ between $O_x$ and $O_y$ in a disordered $d$-symmetry FF density wave.

B. Measured $Z(r,150\text{meV})$ at $T > T_c$ in the pseudogap phase of hole density $p \sim 8\%$ doped $Bi_2Sr_2CaCu_2O_8$ ($T \sim 45K$). Although correlation lengths are shorter, the dFF-DW phenomena are otherwise indistinguishable from observations at $T \ll T_c$.

C. The $d$-symmetry form factor power spectral density $|\tilde{D}^Z(q)|^2 = |(\tilde{O}_x(q) - \tilde{O}_y(q))/2|^2$ determined from sublattice-phase resolved analysis of data in A. Two primary DW peaks at $Q_x$ and $Q_y$ exist with this $d$-symmetry form factor, as identified by dashed circles.

D. The $d$-symmetry FF power spectral density $|\tilde{D}^Z(q)|^2 = |(\tilde{O}_x(q) - \tilde{O}_y(q))/2|^2$ determined from sublattice-phase resolved analysis of data in B. Again, two primary DW peaks at $Q_x$ and $Q_y$ exist with this $d$-symmetry form factor, showing that the $q$-space structure of dFF-DW phenomenology is identical in the pseudogap phase and in the superconducting phase.

E. F. Using only the regions within the dashed circles in C,D, the $r$-space amplitudes of the dFF-DW in A,B are calculated for modulations along $Q_x$ from Eqn. 2a, and along $Q_y$ from Eqn. 2b. Then using $F(r) = (|D_x(r)| - |D_y(r)|)/(|D_x(r)| + |D_y(r)|)$ (see SI Section 3) regions primarily modulating along $y$-axis with $-1.0 < F(r) < -0.3$ are shaded blue. Regions primarily modulating along $x$-axis with $+0.3 < F(r) < +1.0$ are shaded orange

F. Domain configuration of unidirectional dFF-DW modulations contained in Fig. 1A at $T \ll T_c$. The unidirectionality color scale for $F(r)$ demonstrated in E,F is overlaid upon the data in A. The dashed circle shows the $r$-space radius equivalent to the $q$-space filter used to generate the $Dx,y(r)$ images by Fourier filtering (see SI Section 4).

G. Domain configuration of unidirectional dFF-DW modulations contained in Fig. 1B at $T>T_c$. The unidirectionality color scale for $F(r)$ demonstrated in E,F is overlaid upon the data in B. The dashed circle has same definition as in G.

**Figure 3 Concentration of dFF-DW Spectral Weight on Pseudogap Energy**

A. The tunneling density of states spectrum $g(E=eV)=dI/dV(E)$ typical of underdoped cuprates show for the $p\sim8\%$ samples presented in this paper. The energy $\Delta_0$ beyond which Bogoliubov QPI do not exist[32,33] and the pseudogap energy $\Delta_1$ are indicated.

B. The energy dispersion of seven dispersive modulation characteristic of Bogoliubov quasiparticle excitations of a $d$-wave superconductor ($q_1...q_7$). These Bogoliubov quasiparticle interference modulations are all simultaneously observable only below the energy $\Delta_0$ as indicated by dashed red line (Refs.31, 32, 33); here we demonstrate that they exhibit a predominantly $s'$-symmetry form factor indicated by the red color. At energies above $\Delta_0$, the electronic structure images evolve quickly to consist of only non-dispersive $Q_x^d$ and $Q_y^d$ wavevectors of the $d$-symmetry form factor DW. We plot the dispersion of these modulations as the energy dependence of the maxima in $D(q,E)$ using blue squares. The same physical modulations when analyzed using $S'(q,E)$ appear as the non-dispersive $Q_x^{s'}$ and $Q_y^{s'}$ wavevectors shown as red circles.

C. Measured $|\widetilde{D}^Z(q, 88 \text{ meV})|^2$ for samples studied herein. The $Q_x^d$ and $Q_y^d$ wavevectors of the $d$-symmetry form factor DW are indicated by two red circles; the data of relevance for determining energy/momentum dependence of the dFF DW modulations is contained within.

D. Measured doping dependence of $Q_x$ and $Q_y$ of $d$-symmetry form factor DW in underdoped $Bi_2Sr_2CaCu_2O_{8+x}$ is shown using blue symbols. The measured doping dependence of $q$-vector linking tips of arcs beyond which the signature of Bogoliubov quasiparticles disappears (E & Refs.31,32,33) shown by all other colors.

E. Measured doping dependence of the wavevectors interconnecting the **k**-space arc tips at which Bogoliubov quasiparticle signatures disappear (B & Refs.31,32,33).

F. Measured energy dependence of $S'(E)$ where **q** is integrated over the region inside solid red circles in C, is shown in red. Measured energy dependence of $S(E)$ where **q** is integrated over the region inside solid red circles in C, is shown in black. Measured energy dependence of *d*-symmetry form factor $D(E)$ where **q** is integrated over the region inside solid red circles in C is shown in blue (SI Section 4). These data reveal that the dFF-DW spectral weight is concentrated at energy surrounding ~90meV which, at this hole density, is the independently measured pseudogap energy scale $\Delta_1$ (see Fig. 3A) and indicated on all relevant panels by a dashed blue line.

**Figure 4 Relationship between dFF-DW modulations of filled and empty states**

A. Differential tunneling conductance image *g*(**r**,+94meV) measured above $E_F$ near the pseudogap energy +$\Delta_1$. The color scale is reversed compared to B.

B. Differential tunneling conductance image *g*(**r**,-94meV) measured below $E_F$ near the pseudogap energy -$\Delta_1$. The color scale is reversed compared to A.

C. The spatial phase of the dFF-DW modulating along the *y*-direction $\phi_y(r,E)$ is calculated using Eqn. 4a from *g*(**r**,+94meV) data in A. The dashed circle shows the *r*-space radius equivalent to the *q*-space filter used to generate the $\phi_{x,y}(r,E)$ images by Fourier filtering (see SI Section 6).

D. The spatial phase of the dFF-DW modulating along the *y*-direction $\phi_y(r,E)$ is calculated using Eqn. 4b from *g*(**r**,-94meV) data in B.

E. & F. From the FOV of A and B, we show the energy dependence of the relative phase of *g*(**r**,-E) and *g*(**r**,+E) modulations along the *y*-axis (SI Section 5): $\phi_y(r,+E) - \phi_y(r,-E)$ when averaged over every pair of identical pixel locations *r*; similarly for relative phase of *g(r,-E)* and *g(r,+E)* for modulations along the *x*-axis: $\phi_x(r,+E) - \phi_x(r,-E)$. The low energy $E<\Delta_0$ Bogoliubov quasiparticle modulations at *+E* and *-E* are in phase spatially and so have relative phase difference of 0. As the pseudogap energy $\Delta_1$ is approached and the dFF-DW phenomena emerge, the relative spatial phase of empty-state an filled state dFF-DW modulations varies wildly in the narrow energy grange shaded gray, and the quickly develops a robust phase shift of π.

**Acknowledgements**

We acknowledge and thank H. Alloul, D. Chowdhury, R. Comin, A. Damascelli, E. Fradkin, D. Hawthorn, J. E. Hoffman, M.-H. Julien, D. H. Lee, M. Norman, and C. Pepin for helpful discussions and communications. We are especially grateful to S.A. Kivelson for key scientific discussions and advice. Experimental studies were supported by the Center for Emergent Superconductivity, an Energy Frontier Research Center, headquartered at Brookhaven National Laboratory and funded by the U.S. Department of Energy under DE-2009-BNL-PM015, as well as by a Grant-in-Aid for Scientific Research from the Ministry of Science and Education (Japan) and the Global Centers of Excellence Program for Japan Society for the Promotion of Science. CKK acknowledges support under *FlucTeam* program at Brookhaven National Laboratory Contract DE-AC02-98CH10886. S.D.E., J.C.D. and A.P.M acknowledge the support of EPSRC through the Programme Grant '*Topological Protection and Non-Equilibrium States in Correlated Electron Systems*". Theoretical studies at Cornell University were supported by NSF Grant DMR-0955822. Theoretical studies at Harvard University were supported by NSF Grant DMR-1103860 and by the Templeton Foundation. Research at Perimeter Institute is supported by the Government of Canada through Industry Canada and by the Province of Ontario through the Ministry of Research and Innovation.

Figure 1

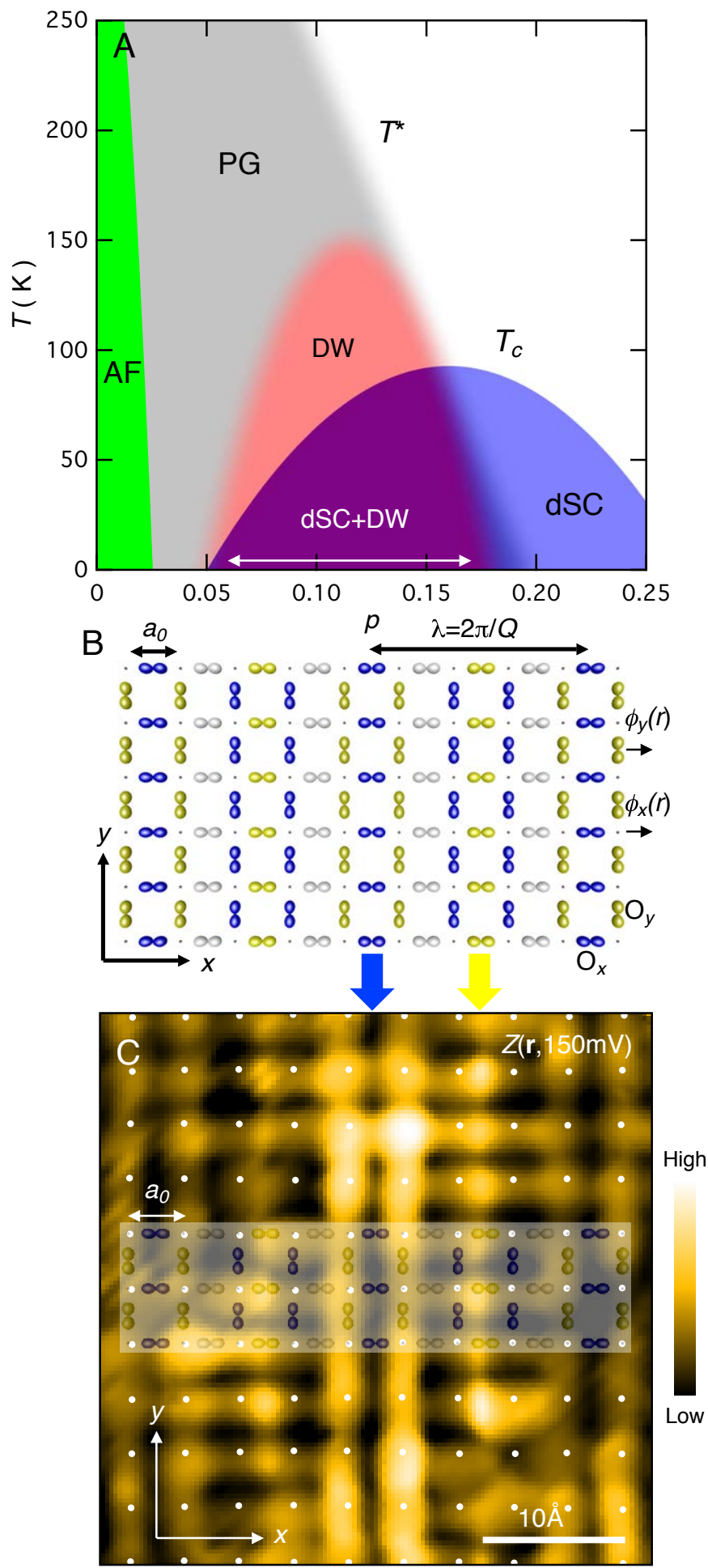

Figure 2

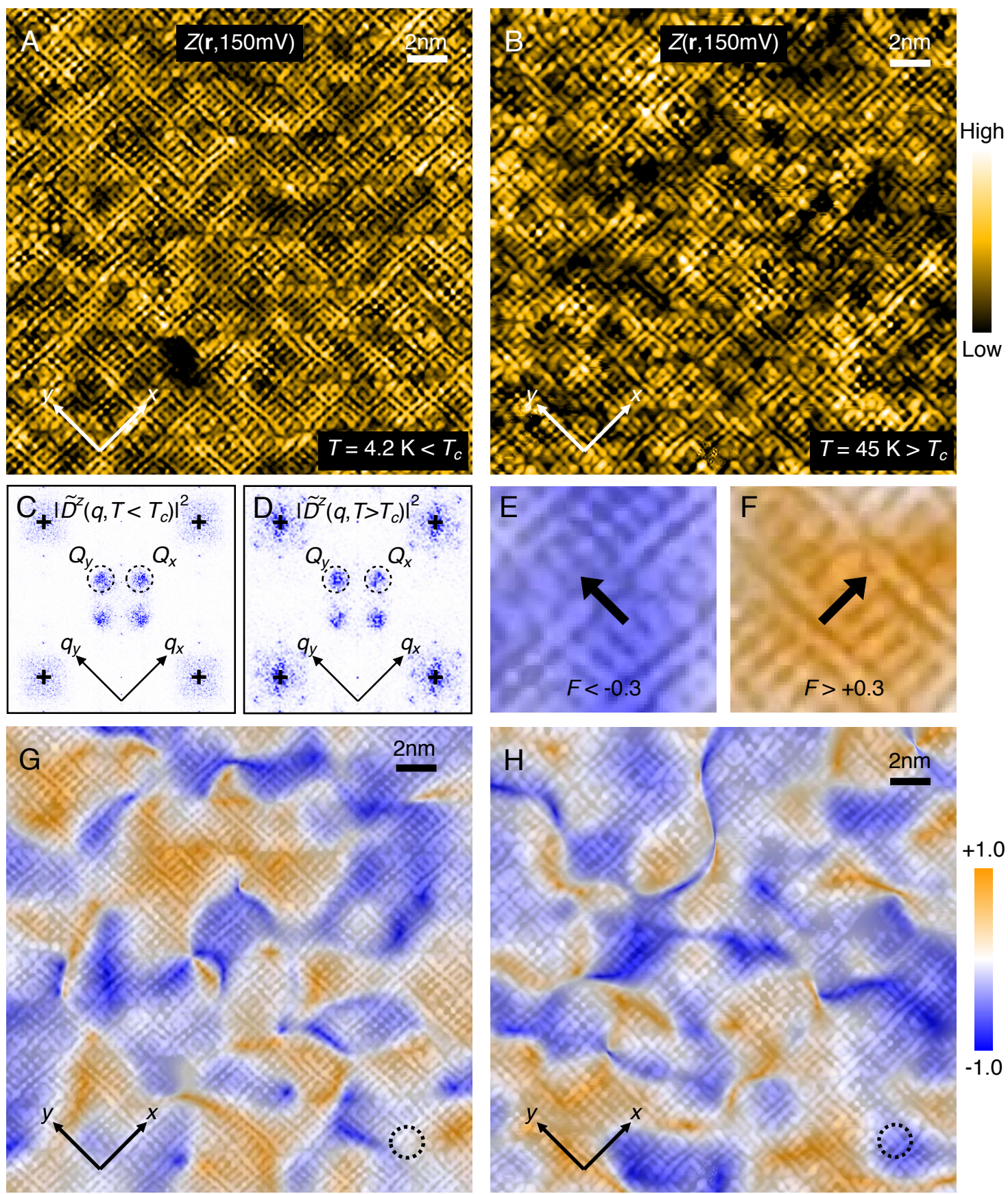

Figure 3

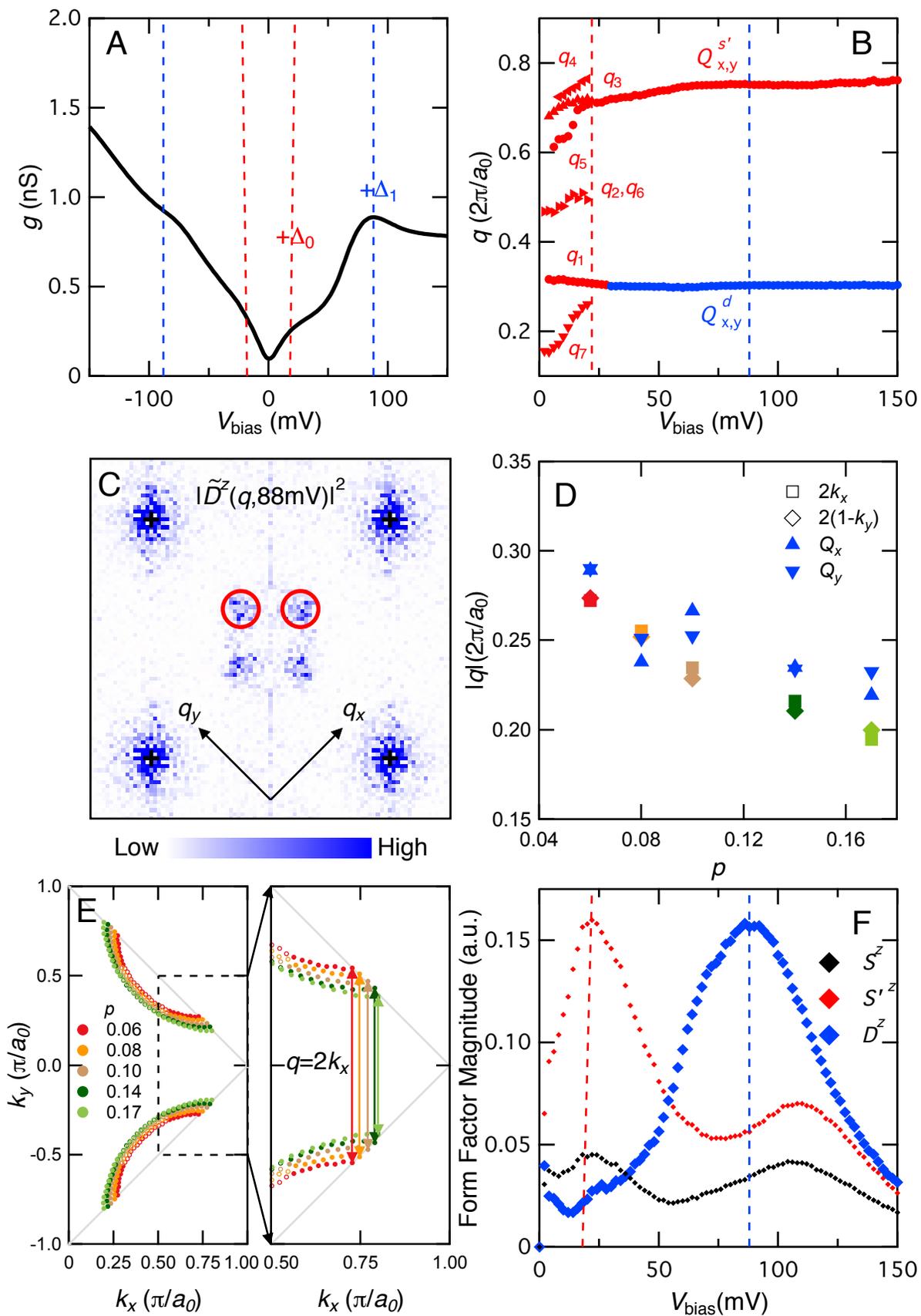

Figure 4

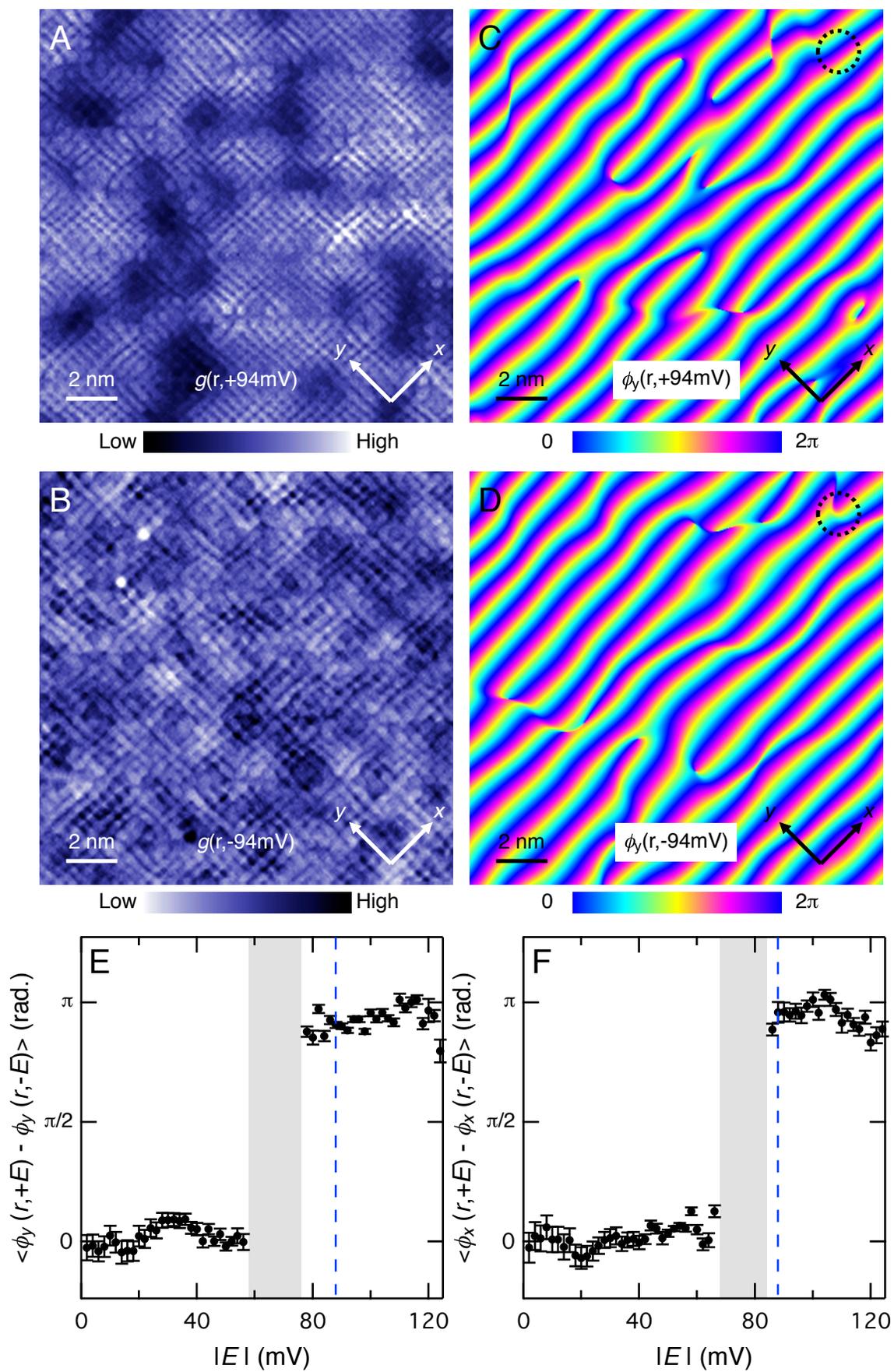

Supplementary Information

# Atomic-scale Electronic Structure of the Cuprate
# *d*-Symmetry Form Factor Density Wave State

M. H. Hamidian, S.D. Edkins, Chung Koo Kim, J. C. Davis, A. P. Mackenzie,
H. Eisaki, S. Uchida, M. J. Lawler, E.-A. Kim, S. Sachdev, and K. Fujita## SI 1. Form Factor Decomposition of CuO₂ IUC States

Here we present mathematical details behind the form factor organization of density waves on the CuO$_2$ plane. The different density wave form factors are due to periodic modulation of the $Q = 0$ form factors, whose point group symmetry is well defined. Modulating these form factors with wave-vector $Q \neq 0$, one obtains

$$A_S(r) = \begin{cases} S\cos(Q \cdot r + \phi_S), & r = r_{Cu}, \\ 0, & r = r_{O_x}, \\ 0, & r = r_{O_y}, \end{cases} \quad A_{S'} = \begin{cases} 0, & r = r_{Cu}, \\ S'\cos(Q \cdot r + \phi_{S'}), & r = r_{O_x}, \\ S'\cos(Q \cdot r + \phi_{S'}), & r = r_{O_y}, \end{cases}$$

$$A_D(r) = \begin{cases} 0, & r = r_{Cu}, \\ D\cos(Q \cdot r + \phi_D), & r = r_{O_x}, \\ -D\cos(Q \cdot r + \phi_D), & r = r_{O_y}, \end{cases} \quad (\text{S1.1})$$

In the CuO$_2$ plane of cuprates, $r_{Cu}$, $r_{O_x}$ and $r_{O_y}$ are the Cu, O$_x$ and O$_y$ sublattice sites, and $\phi_{S,S',D}$ are the phases of each of the density wave form factor components. Equation S1.1 shows that a purely *d*-symmetry form factor density wave can be thought of as a wave on each of the oxygen sub-lattices but with a spatial phase shift of π between them. The Fourier transform of the *d*-symmetry form factor density wave is presented in Figs. S2 and will be considered further in SI section 3

# SI 2: Setup Effect in SI-STM Measurements

Spectroscopic imaging scanning tunneling microscopy (SI-STM) measurements provide energy dependent electronic structure images that can be used to identify the presence of modulations whether they arise from dispersive quasi-particle interference (QPI) or ultra-slow-dispersive density waves (DW) – see K. Fujita et al *Strongly Correlated Systems - Experimental Techniques* by Springer (ISBN 978-3-662-44132-9). However, the protocol by which a tunneling junction is established in SI-STM measurements can transfer conductance modulations from one set of energies to another. This systematic error which results in a misidentification of the energy of states undergoing spatial modulations is called the *setup effect*. Therefore, determining the physically real modulations and especially their correct energy can present a grave challenge. Ignoring the setup effect, as is often the case, leads to incorrect characterization of electronic structure properties of materials.

## I - *Mathematical Description of STM Observables*

The basic observable in STM experiments is the tunneling current, *I*, which depends on the bias between the tip and the sample, *V*, the tip sample separation, *z*, and the position along the sample, **r**:

$$I(\mathbf{r}, z, V) = f(\mathbf{r}, z) \int_0^{eV} LDOS(\mathbf{r}, \epsilon) d\epsilon \quad (S2.1)$$

The function $f(\mathbf{r}, z)$ captures spatial variations due to surface corrugation, work function, matrix elements and proximity *z* of the tip to the surface. The integral of the local density of states, *LDOS*, includes spatial variation of the electronic structure.

Spectroscopic imaging entails establishing a stable tunnel junction at every *r* by using the same arbitrary pre-chosen set points $I_S$ and $V_S$, and then measuring the variation with *r* of the current or of the differential conductance, at each bias *V*. The set point constraints then determine the pre-factor of the integral in equation (S2.1):

$$I(\mathbf{r}, z, V_s) = I_s = f(\mathbf{r}, z) \int_0^{eV_s} LDOS(\mathbf{r}, \epsilon) d\epsilon$$

$$\Rightarrow f(\mathbf{r}, z) = \frac{I_s}{\int_0^{eV_s} LDOS(\mathbf{r}, \epsilon) d\epsilon}$$

and thus the expressions for the spectroscopic current and differential conductance *dI/dV* are given by

$$I(\mathbf{r}, V) \propto \frac{\int_0^{eV} LDOS(\mathbf{r},\epsilon) d\epsilon}{\int_0^{eV_s} LDOS(\mathbf{r},\epsilon) d\epsilon} \quad , \quad \frac{dI}{dV}(\mathbf{r}, V) \propto \frac{LDOS(\mathbf{r},eV)}{\int_0^{eV_s} LDOS(\mathbf{r},\epsilon) d\epsilon} \quad , \quad (S2.2)$$

The term in the denominator is responsible for the deleterious nature of the setup effect since it carries the imprint of electronic structure over the whole range of voltages between 0 and the set-up bias, $V_S$.

**II - *Example of the Setup Effect***

The figure below demonstrates how the choice of setup bias to establish the tunneling junction strongly influences the acquired data. While both the left and right panels are spatial images of the differential conductance taken at *V* = 50mV in the same field of view, the left was measured with $V_s$ =150mV setup bias while the right with $V_s$ = 50mV. It is clear that the spatial intensity patterns of the same set of states in the material are imaged differently based on the setup bias parameter.

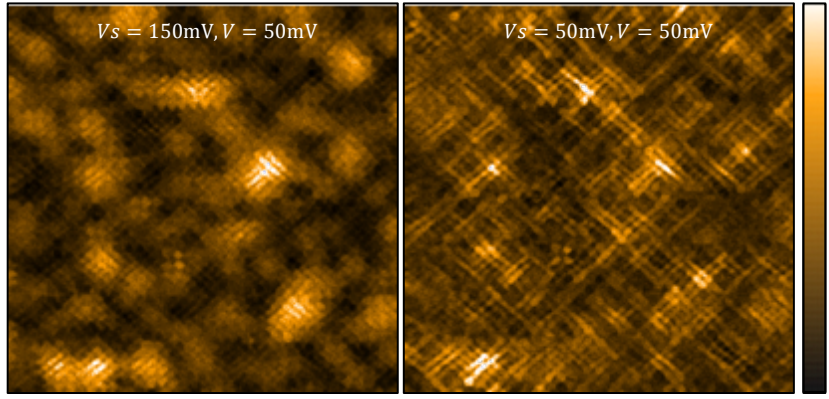

**Figure S1: Example of Setup Effect** Two panels show the spatial differential conductance of underoped BSCCO-2212 at 50mV acquired in the same field of view. The left panel was acquired with a setup bias of 150mV while the right with one of 50mV.

We emphasize that the situation shown in Fig. S1 is not an unusual effect but rather is completely typical and unavoidable in all SI-STM studies, except under some specially prepared conditions (Section 6).

## SI 3. Predicted Fourier Transform STM Signatures of a dFF-DW

**I –** *Sublattice Segregation Method to Determine DW Form Factor Symmetry*

This section predicts the consequences of a primarily *d*-symmetry form factor density wave for the complex Fourier transform images of electronic structure on the three sublattices: $\mathrm{Cu}(\boldsymbol{q})$, $\tilde{O}_x(\boldsymbol{q}), \tilde{O}_y(\boldsymbol{q})$. In turn this controls the three key functions used to measure modulations with different form factor symmetry as discussed in the main text.

$$\tilde{S}(\boldsymbol{q}) = \widetilde{Cu}(\boldsymbol{q}) \quad (S3.1)$$

$$\tilde{S}'(\boldsymbol{q}) = \left(\tilde{O}_x(\boldsymbol{q}) + \tilde{O}_y(\boldsymbol{q})\right)/2 \quad (S3.2)$$

$$\tilde{D}(\boldsymbol{q}) = \left(\tilde{O}_x(\boldsymbol{q}) - \tilde{O}_y(\boldsymbol{q})\right)/2 \quad (S3.3)$$

A purely *d*-symmetry form factor density wave has modulations in anti-phase on the x and y oxygen sub-lattices (Fig. 1B of main text) and no modulation on the Cu site. For the specific example of $\boldsymbol{Q_x} \approx \left(\frac{1}{4}, 0\right)$, $\boldsymbol{Q_y} \approx \left(0, \frac{1}{4}\right)$ considered in our study, this requires that the peaks at $\pm\boldsymbol{Q_x}$ and $\pm\boldsymbol{Q_y}$ present in both $\widetilde{O}_x(\boldsymbol{q})$ and $\widetilde{O}_y(\boldsymbol{q})$ must cancel exactly in $\widetilde{S}'(\boldsymbol{q}) = \left(\widetilde{O}_x(\boldsymbol{q}) + \widetilde{O}_y(\boldsymbol{q})\right)/2$ and be enhanced in $\widetilde{D}(\boldsymbol{q}) = \left(\widetilde{O}_x(\boldsymbol{q}) - \widetilde{O}_y(\boldsymbol{q})\right)/2$. Conversely the peaks at $\boldsymbol{Q'}$ = (1,0)±$\boldsymbol{Q_{x,y}}$ and $\boldsymbol{Q''}$=(0,1)±$\boldsymbol{Q_{x,y}}$ will be enhanced in $\widetilde{S}'(\boldsymbol{q})$ but will cancel exactly in $\widetilde{D}(\boldsymbol{q})$ (Figs S2B,C).

This occurs because the two sub-lattices have modulations at the same wave-vector but with a π phase shift between them. Importantly, electronic structure images formed using the difference of oxygen sub-lattices, as in $\widetilde{D}(\boldsymbol{q})$, have the effect of removing this phase difference and recovering the peaks in the Fourier transform at the fundamental wave-vectors. These are necessary consequences of a density wave with a primarily *d*-symmetry form factor and hold for any *d*-symmetry form factor modulation in the presence of arbitrary amplitude and overall phase disorder.

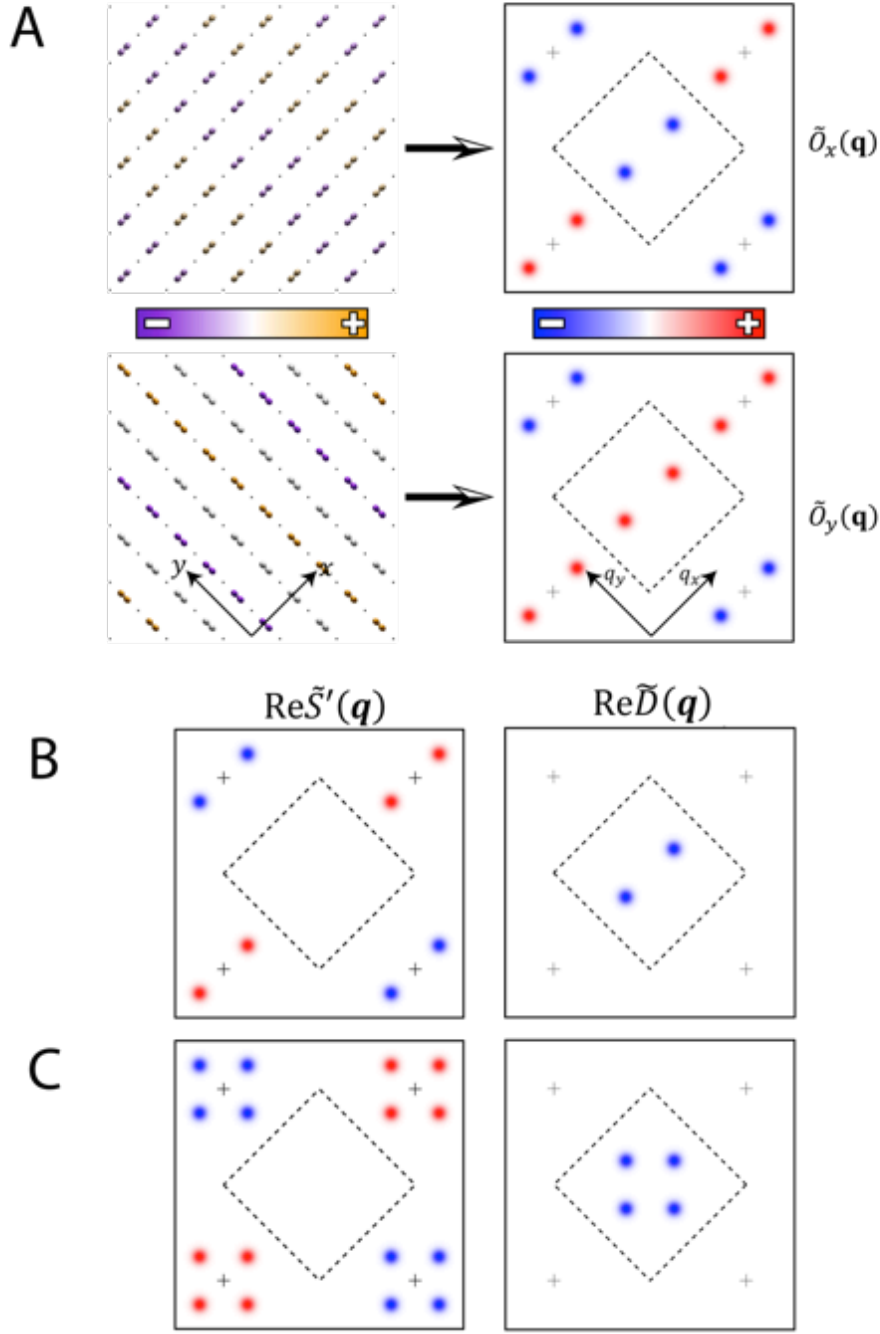

**Figure S2: Fourier Analysis of the Density Wave** (A) Schematic of the segregated sublattice images $O_{x,y}(r)$ and their Fourier transforms $\tilde{O}_{x,y}(q)$ (B) $Re\tilde{S}'(q)$ and $Re\tilde{D}(q)$ for a *d*-form factor density wave with modulation along the x direction at $Q=(Q,0)$. Note that the origin of co-ordinates in real space has been chosen such that the Fourier transforms are purely real. (C) $Re\tilde{S}'(q)$ and $Re\tilde{D}(q)$ for a d-form factor density wave with modulations along the x and y directions at $Q=(Q,0),(0,Q)$. The key signature of the d-form factor is the absence of the peaks at $(Q,0),(0,Q)$ in $Re\tilde{S}'(q)$ and their presence in $Re\tilde{D}(q)$; the converse being true for the DW peaks surrounding $(\pm 1, 0)$ and $(0, \pm 1)$.

## II – Form factor symmetry conversion of $q_5$ for $E > \Delta_0$

Fig. 3B of the main text shows the quasiparticle interference scattering vectors $q_1$-$q_7$ for energies below $\Delta_0$ which are primarily s'FF symmetry modulations in the Z-map. Above this energy scale the dFF modulation appears, plotted in blue at $Q_{x,y}^d$. This low-$q$ modulation is extracted according to the scheme presented above by which the sublattice subtraction procedure of $O_x$ and $O_y$ reveals the density wave at its primary wavevector. However, as seen from Fig. S2 B,C above, the physically identical dFF-DW can also be seen in the $O_x(q)+O_y(q)$ at a $q$-vector which is $Q_{x,y}^{s'} = 1 - Q_{x,y}^d$. Thus, the same physical modulation will manifest at different but related $q$-vectors in S'($q$) and D($q$). Figure S3 demonstrates this dual representation of the same phenomenon. In Fig. 3B, the dFF-DW peak in S'($q$) is plotted as $Q_{x,y}^{s'}$ while that in D($q$) is plotted as $Q_{x,y}^d$; they represent the same physical dFF-DW modulation. In S'($q$) the dFF-DW modulation appears as $Q_{x,y}^{s'} \sim$ q$_5(\Delta_0)$ persisting to energies much larger than $\Delta_0$. In D($q$) the identical physical modulation appears in the same energy range at a wavevector $Q_{x,y}^d \sim$ q$_1(\Delta_0)$ (Fig. 3F).

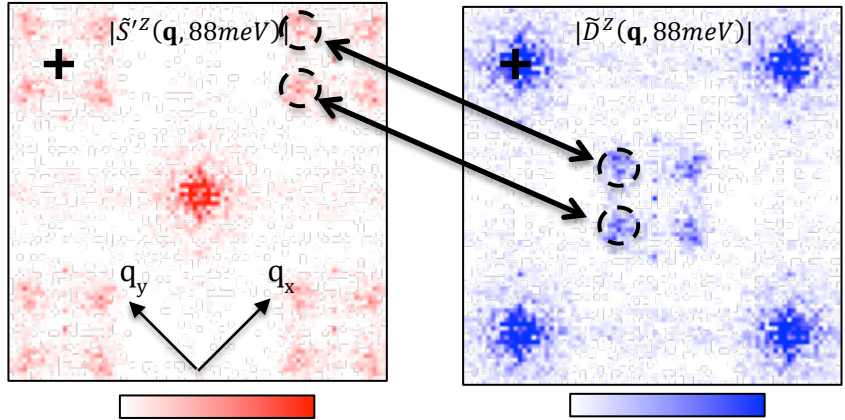

**Figure S3: Equivalent representation of dFF modulations as $Q^{s'}$ in the S' and $Q^d$ in the D channels.**

## SI 4: Calculation of the local nematicity *F(r)*

We define the local nematicity of the DW to be the function

$$F(\boldsymbol{r}) = \frac{|D_x(\boldsymbol{r})| - |D_y(\boldsymbol{r})|}{|D_x(\boldsymbol{r})| + |D_y(\boldsymbol{r})|} \tag{S4.1}$$

where $|D_x(\boldsymbol{r})|$ and $|D_y(\boldsymbol{r})|$ are the amplitudes of the *d*-symmetry form factor components of the density wave at wave-vectors $\boldsymbol{Q}_x \approx \left(\frac{1}{4}, 0\right)$ and $\boldsymbol{Q}_y \approx \left(0, \frac{1}{4}\right)$ respectively. Hence, in order to calculate this function we must first experimentally determine $|D_{x,y}(\boldsymbol{r})|$. This may be achieved through Fourier filtration of the measured $\widetilde{D}(\boldsymbol{q}) = (\widetilde{O}_x(\boldsymbol{q}) - \widetilde{O}_y(\boldsymbol{q}))/2$ to select the amplitude contained in the region of reciprocal space proximately surrounding $\boldsymbol{Q}_x$ and $\boldsymbol{Q}_y$. The function $\widetilde{D}(\boldsymbol{q})$ is multiplied by a gaussian of FWHM $2\sqrt{2\ln 2}\Lambda$ centred on $\boldsymbol{Q}_x$ and $\boldsymbol{Q}_y$ respectively to create two filtered complex Fourier transforms $D_x(\boldsymbol{q})$ and $D_y(\boldsymbol{q})$.

$$\widetilde{D}_{x,y}(\boldsymbol{q}) = 2\widetilde{D}(\boldsymbol{q}) e^{-\frac{(q - Q_{x,y})^2}{2\Lambda^2}} \tag{S4.2}$$

Here, the factor of two arises from the fact that we have only filtered around $\boldsymbol{Q}_{x,y}$ and not $-\boldsymbol{Q}_{x,y}$ which, by the inversion symmetry of the Fourier transform of real functions, contain identical information. By taking the modulus of their inverse Fourier transforms one obtains an estimate of the amplitude of the x and y directed modulations.

$$|D_{x,y}(\boldsymbol{r})| = \left| \frac{1}{(2\pi)^2} \int d\boldsymbol{q}\, e^{i\boldsymbol{q}\cdot\boldsymbol{r}} \widetilde{D}_{x,y}(\boldsymbol{q}) \right| \tag{S4.3}$$

# SI 5: Cuprate octet scattering QPI, calculation of energy resolved form factor magnitudes, and results for higher doped samples

*I – The octet scattering vectors of Bogoliubov quasi-particle interference*

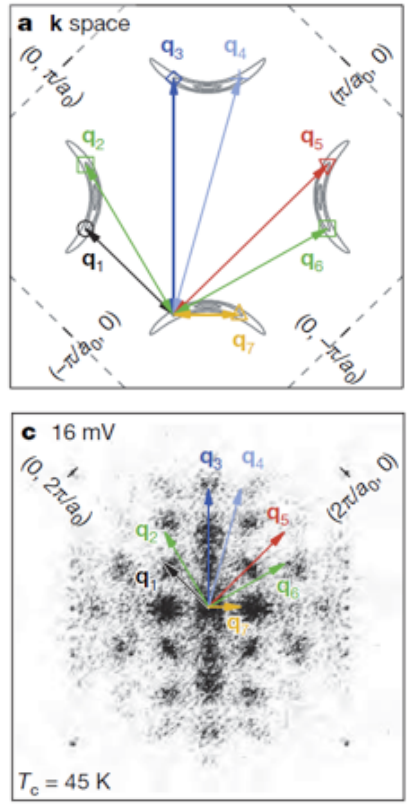

**Fig S4: Cuprate Octet Model for QPI** Top panel shows the constant energy contours of the superconducting cuprate **k**-space electronic structure. The tips of the arc are connected via a set of 7 scattering vectors $q_1$-$q_7$. The bottom panel shows the dominant scattering vectors in **q**-space detected from quaisparticle interference imaging (QPI) in spectroscopic STM imaging. The scattering vectors can be used to reconstruct the energy resolved **k**-space electronic structure.

*II – Spectroscopic determination of form factor magnitudes*

The main texts presents the energy resolved weight of various form factor modulations in the spectroscopic data in Figure 3F. This section describes the set of operations to generate such curves.

First, sub-lattice segregation of the measured $Z(r,E)$ data is implemented by extracting the intensities at Cu, $O_x$, and $O_y$ sites from which, respectively, one constructs the energy resolved form factor images $S(r, E), S'(r, E)$, and $D(r, E)$, as prescribed in SI Section 3.

From these real space images one generates the power spectral density for each of the form factors, $|\tilde{S}^Z(q, E)|^2$, $|\tilde{S}'^Z(q, E)|^2$, and $|\tilde{D}^Z(q, E)|^2$. Fig. S3 presents a subset of these Fourier transform images taken at $E$ = 90meV. Finally, integration is carried out over $q$–space regions inside the broken circles at each energy to obtain the spectral signature for each form factor, as plotted in Fig. 3F of the main text.

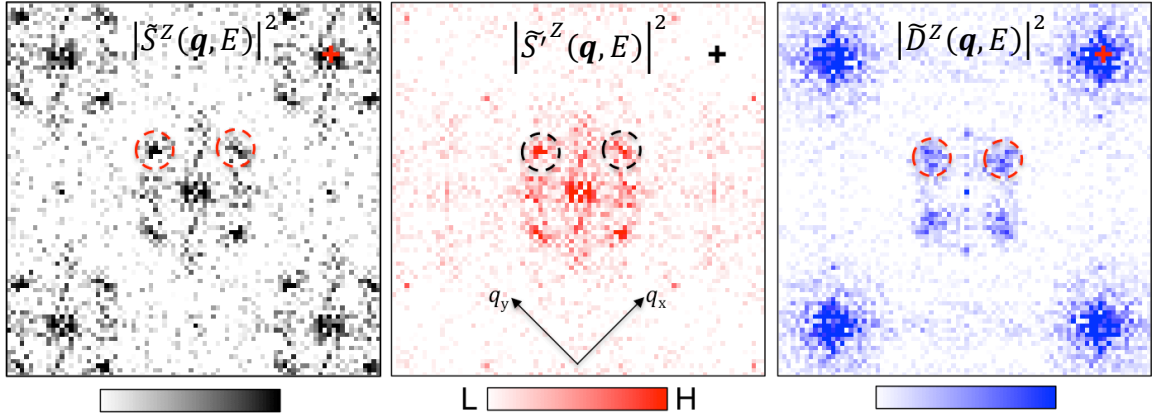

**Figure S5: Spectral Densities for Form Factors** Power spectral density for S(A), S'(B), and D(C) symmetry form factor at representative energy. Bragg peak locations are marked by "+". $q$-space regions inside broken circles are the region of interest to calculate the spectral weight for each form factor.

## SI 6: DW Phase Shift Between Empty and Filled States

Establishing the actual existence of a physical density wave at any given energy is the first step to determining the phase difference between modulations in the empty and filed states. In light of the setup effect, as introduced in SI Section 2, one must proceed carefully and not mistake the imprints of modulations in place of real ones. To avoid such errors one must use a combination of spectroscopic and topographic imaging.

**I – *Using Topographic Data to Determine Setup Bias to Avoid Setup Effect.***

In constant conductance imaging, or *topographic imaging*, the STM feedback system adjusts the tip sample separation, $z$, as it scans over the sample surface to maintain a set point current, $I_S$, at a constant applied tip-sample bias $V_S$. The topographic image, or $z(\mathbf{r})$, can be derived by starting with the equation for the tunneling current,

$$I(\mathbf{r}, z, V) = f(\mathbf{r}, z) \int_0^{eV} LDOS(\mathbf{r}, \epsilon) d\epsilon \qquad (S6.1)$$

and assuming that the function $f(\mathbf{r}, z)$, which represents the effect of corrugation, work function, and tunneling matrix elements, takes the form

$$f(\mathbf{r}, z) = \exp(-\kappa z) \cdot A(\mathbf{r}) \qquad (S6.2)$$

where $\kappa$ depends on a mixture of the work functions of the sample and tip. Then for topographic imaging the recorded value of the relative tip-sample separation takes the form

$$z(\mathbf{r}) = \frac{1}{\kappa} \ln \left[ \int_0^{eV_S} LDOS(\mathbf{r}, \epsilon) d\epsilon \right] + \frac{1}{\kappa} \ln \left[ \frac{I_S}{A(\mathbf{r})} \right]. \qquad (S6.3)$$

The essential point is that a high signal-to-noise topographic image obtained by STM in constant conductance mode reveals contributions from both the surface structure and variations in the $LDOS(\mathbf{r}, E)$ according to S6.3, obviously provided $E$ <$eV_s$. In particular, through this effect a density wave imprints its signature logarithmically onto a topographic image beyond what would be expected from a slight deformation of the lattice due to electron-lattice coupling.

By choosing a setup bias such that the denominator in equation S2.2 is a constant without any periodic structure then any modulations observed in the differential tunneling conductance images, $g(\mathbf{r}, E)$, must be representative of real modulations in the *LDOS*. As a matter of practicality, however, such a set up condition is highly unlikely and it is only possible to choose $V_s$ for which only a subset of modulations are not manifest. For the purposes of analyzing the cuprate density we have chosen $V_s$ so that the dFF-DW has no spectral weigh in the setup topographs and thus no weight in the denominator of S2.2 (see fig. S6 below).

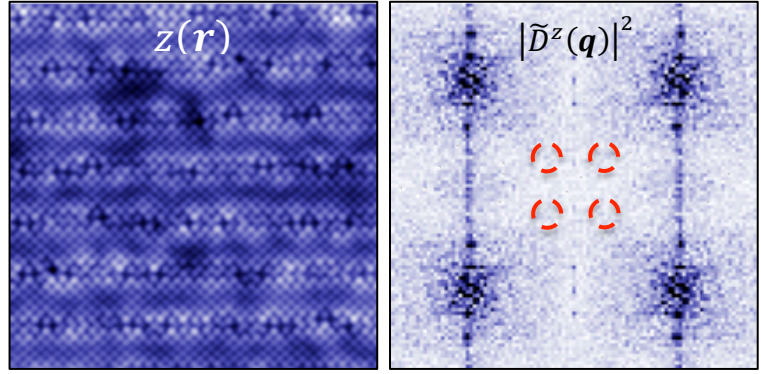

**Figure S6: Using topography to Avoid Setup Effect** (Left panel) Topographic image of UD-BSCCO acquired at a setup bias of 250mV. (Right panel $|\tilde{D}^z(\mathbf{q})|^2$ which measures the components of dFF-DW in the topographic image. No peaks are observed inside the dashed red circles enclosing $Q_{DW}$.

From topography taken at different applied bias values (see equation S6.1), one can determine which bias produces a topographic image that has no dFF-DW signal arising from spatial modulations in the *LDOS*. The left panel of Fig. S6 presents a topographic image taken on UD-BSCCO at $V_s$ = 250mV while the right panel shows its corresponding $|\tilde{D}(\mathbf{q})|^2$, used to determine if there exists a dFF-DW component at $Q_{DW}$. There are no peaks at $Q_{DW}$ meaning that a spectroscopic setup bias of $V_s$ = 250mV for measuring $g(\mathbf{r}, E)$ will not lead to spurious imprints of dFF-DW modulation on the acquired $g(\mathbf{r}, E)$ data. Hence, all dFF-DW modulations observed at any energy can be interpreted as physically real.

The main text presents $g(\mathbf{r}, E)$ data for underdoped BSCCO. Because the data was acquired with the setup condition $V_s$ = 250mV all dFF-DW modulations in all of the energy layers, both on the empty and filled sides, can be treated as physically real modulations and not due to the setup effect of Section 2. It is then meaningful to compare the phases of the dFF modulations between energy layers. With this justification Fig. 4E of main text presents the phase difference between $g(\mathbf{r}, E)$ and $g(\mathbf{r}, -E)$ without the setup effect error .

## II  *Setup-Bias-Dependent DW Phase Shift*

As described above, the correct choice of set up bias is essential in determining the existence of physically real density wave modulations on both the filled and empty states of the electronic structure. By choosing a bias for which the integrated density of states has no signature of the dFF-DW, it is ensured that there is no systematic error in the differential conductance data with a false dFF-DW signal. Such a requirement led us to search for and find a specific setup bias value $Vs > \Delta_1$ in order get a measure of the true phase shift between the empty and filled state density wave modulations (see SI Section V). However only a small range of such setup bias conditions with no topographic modulation at $\mathbf{Q}$ plus which allowed for stable experimental conditions was achievable, meaning that repeating this procedure at several setup conditions proved impossible

Nevertheless, to further demonstrate that the π phase shift between empty/filled dFF-DW modulations cannot be due to the setup effect, we plot the value of this phase shift at $|E| = \Delta_1$ for a number of arbitrary $Vs$. If the phase difference were caused merely by the setup effect then the phase shift would always be $\pi$. But, as the plot clearly demonstrates, the phase difference is dependent on the setup bias with the lower choices for $V_s$ producing arbitrary shifts due to the setup effect, as expected.

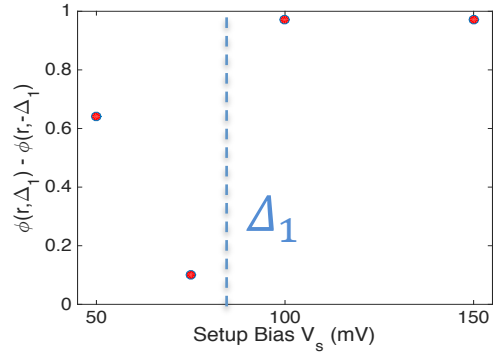

**Figure S7:** Phase shift between the empty and filled side density wave modulations at energy $\Delta_1$ for various spectroscopic setup bias values $V_s$. The data is shown for BSCCO-2212 UD45 for which $\Delta_1 \approx 80 meV$.

### III– *Phase Shift Analysis for Multiple Samples at Different Doping*

Figure S8 shows the phase difference between the dFF-DW modulation at E=±$\Delta_1$ in the differential conductance map for higher doped samples. As in Figs. 4E,F of the main text where the |E|∼ $\Delta_1$ density wave acquires a $\pi$ phase shift between empty and filled states for *p*∼6%; here we show highly consistent effects for two other dopings so that the phenomenology actually spans the range 0.06<*p*<0.17.

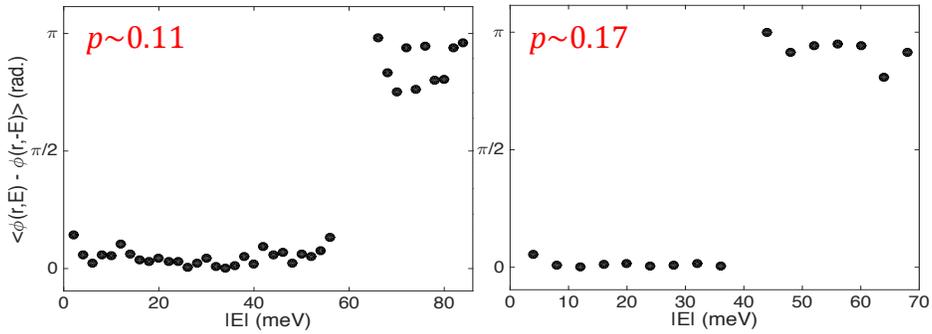

**FIG S8: Phase Shift in UD samples** dFF-DW phase shift between empty and filled for *p*∼ 0.1 and *p*∼0.17.